\documentclass{llncs}
\usepackage{graphicx}
\usepackage{subcaption}
\usepackage{enumitem}
\usepackage{color}
\usepackage{algorithm}
\usepackage{algorithmicx}
\usepackage[noend]{algpseudocode}
\usepackage{amsmath}

\graphicspath{ {./Images/} }

\begin{document}

\title{Computing a Feedback Arc Set Using PageRank}

\titlerunning{Computing a Feedback Arc Set Using PageRank}
%\title{Feedback Arc Set Revisited: An Approach Using PageRank}

%\title{Feedback Arc Set Using PageRank}

%\title{Feedback Arc Set: An Approach Using PageRank}

% If the paper title is too long for the running head, you can set
% an abbreviated paper title here
%
%\iffalse
%\author{V. Geladaris \and P. Lionakis \and I. G. Tollis}
\author{Vasileios Geladaris \and Panagiotis Lionakis \and Ioannis G. Tollis}

\authorrunning{V. Geladaris \and P. Lionakis \and I. G. Tollis}
% First names are abbreviated in the running head.
% If there are more than two authors, 'et al.' is used.
%
\institute{Computer Science Department, University of Crete, GREECE \email{{\{csd3926,lionakis,tollis\}@csd.uoc.gr}}
\\}
%\fi
% First names are abbreviated in the running head.
% If there are more than two authors, 'et al.' is used.
%
\maketitle              % typeset the header of the contribution
\begin{abstract}
We present a new heuristic algorithm for computing a minimum Feedback Arc Set in directed graphs. 
The new technique produces solutions that are better than the ones produced by the best previously known
heuristics, often reducing the FAS size by more than 50\%.  It is based on computing the PageRank score of the nodes of the directed line graph of the input directed graph.  Although the time required by our heuristic is heavily influenced by the size of the produced line graph, our experimental results show that it runs very fast even for very large graphs used in graph drawing.
%for visualization. 
%However, it is slower than some previous heuristics for large graphs. 

\end{abstract}
\section{Introduction}
In a directed graph, $G$, a feedback arc set (\emph{FAS}) is a set of edges whose removal leave $G$ acyclic.
The minimum FAS problem is important for visualizing directed graphs in
hierarchical style~\cite{di1999graph}.  In fact, the first step of both known frameworks for
hierarchical graph drawing is to compute a minimum FAS~\cite{JGAA-502,sugiyama1981methods}.
Unfortunately, computing a minimum FAS is NP-hard and thus many heuristics
have been presented in order to find a reasonably good solution. In this paper we
present a new heuristic that uses a different approach and produces FAS that
contain about half the number of edges of the best known heuristics.  However, it requires
superlinear time, and hence it may not be suitable for very large graphs.
Finding a minimum FAS has many additional applications beyond Graph Drawing,
including misinformation removal, label propagation, and many application domains motivated by Social Network Analysis ~\cite{budak2011limiting,he2012influence,simpson2016clearing}.

A feedback arc set of a directed graph $G=(V, E)$ is a subset of edges $F$ of $E$ such that
removing the edges in $F$ from $E$ leaves $G$ acyclic (no directed cycles). In other words,
a FAS contains at least one edge from each cycle of $G$.  In hierarchical drawing algorithms
the edges in a FAS are not removed, but instead their direction is inverted. Following the
terminology of  ~\cite{di1999graph}, a set  of edges whose reversal makes the digraph acyclic is
called a feedback set (FS).  Notice that a FAS is not always a FS.   However, it is easy
to see that every minimal cardinality FAS is also a FS.  Hence it follows that the
minimum FS problem is as hard as the well studied minimum FAS problem which is
known to be NP-hard~\cite{johnson1982np,karp1972reducibility}.
Clearly, any heuristic for solving the minimum FAS problem can be applied for
solving the minimum FS problem, as discussed in~\cite{di1999graph,handbook}.
%~\cite{10.5555/551884}. 

%we need a brief paragraph about previous heuristics...
There have been many heuristics for solving the FAS problem due to the multitude of its applications. Two of the most important heuristics/techniques are
due to Eades, Lin \& Smyth~\cite{eades1993fast} and Brandenburg \& Hanauer~\cite{brandenburg2011sorting}.
The first is a greedy heuristic, that will be called \textit{GreedyFAS}, whereas the second presents a set of heuristics based on sorting.
Simpson, Srinivasan \& Thomo published an experimental study for the FAS problem on very large graphs at web-scale (also called \emph{webgraphs})~\cite{simpson2016efficient}.  They implemented and compared many FAS heuristics. According to their study, the aforementioned are the most efficient heuristics, but only GreedyFAS is suitable to run on their extra large webgraphs.
%An exact method for the minimum feedback arc set problem is recently presented in~\cite{baharev2021} but as expected, even for sparse graphs it takes exponential time.
%that are important for practical applications often have $\Omega (2^n)$ simple cycles

%The first of the two algorithms that currently produce the best FAS size is called $GreedyFAS$ and it is due to Eades, Lin \& Smyth~\cite{eades1993fast}.

In this paper we present a new heuristic algorithm for computing a minimum FAS in directed graphs. 
The new technique produces solutions that are better than the ones produced by the best previous heuristics, sometimes even reducing the FAS size by more than 50\%.  It is based on computing the PageRank score of the nodes of a graph related to the input graph, and runs rather fast for graphs up to 4,000 nodes. However, it is slower than GreedyFAS for webgraphs.

\section{Existing Algorithms}
\label{se:overview}
In this section we summarize and give a brief description of two important heuristics that currently give the best results for the FAS problem, according to the new experimental study of Simpson, Srinivasan \& Thomo~\cite{simpson2016efficient}.
%published an experimental study for the FAS problem 
They implemented and compared many heuristics for FAS, and performed experiments on several large and very large webgraphs.  Their results show that two of the known heuristic algorithms give the best results.

The first of the two heuristic algorithms that currently produce the best FAS size is called \textit{GreedyFAS} and it is due to Eades, Lin \& Smyth~\cite{eades1993fast}. 
%According to~\cite{simpson2016efficient}, a direct implementation of GreedyFAS runs in $O(n^2)$ time but they present two different optimized implementations that run in $O(n + m)$.  
In~\cite{simpson2016efficient} two different optimized implementations of GreedyFAS that run in $O(n + m)$ are presented and tested.  These are the most efficient implementations in their study and are able to run even for their extra large webgraphs.
%of the algorithm has been published by Simpson, Srinivasan \& Thomo that runs in $O(n + m)$ \cite{simpson2016efficient}. 
%and are suitable even for extra large graphs
The second algorithm 
%that we will look at 
is \textit{SortFAS} of Brandenburg \& Hanauer~\cite{brandenburg2011sorting}. According to~\cite{simpson2016efficient}, SortFAS, as proposed runs in $O(n^3)$ time but Simpson et al. present an implementation that runs in $O(n^2)$ time.

We will present experimental results that show that our new heuristic algorithm performs better than both of them in terms of the size of the produced FAS. On the other hand, it takes more time than both of them for large graphs.  However, for graphs that are typically used for visualization purposes, the running time is acceptable whereas the produced FAS size is about half.

%For a comparison of these algorithms to the one we propose in this paper, refer to the related work in Section 4. We will look at the pseudocode for these algorithms and the heuristics they use in order to compute their results.

\subsection{GreedyFAS}
The GreedyFAS algorithm was introduced by Eades, Lin \& Smyth in 1993~\cite{eades1993fast}. It efficiently calculates an approximation to the FAS problem on a graph G. In order to understand the algorithm,  we first discuss the \emph{Linear Arrangement Problem (LA)}, which is an equivalent formulation to the FAS problem. The LA problem produces an ordering of the nodes of a graph $G$ for which the number of arcs pointing backwards is minimum. The set of backwards arcs is a FAS since removing them from $G$ leaves the graph acyclic.

GreedyFAS calculates a feedback arc set of a graph $G$ by first calculating a Linear Arrangement of $G$. More specifically, in each iteration, the algorithm removes all nodes of $G$ that are sinks followed by all the nodes that are sources. It then removes a node $u$ for which $\delta(u) = d^+(u) - d^-(u)$ is a maximum, where $d^+(u)$ denotes the out-degree of $u$ and $d^-(u)$ denotes the in-degree of $u$. The algorithm also makes use of two sequences of nodes $s_1$ and $s_2$. When any node $u$ is removed from $G$ then it is either prepended to $s_2$ if it's a sink, or appended to $s_1$ if it's not. The above steps are repeated until $G$ is left with no nodes, then the sequence $s = s_1s_2$ is returned as a linear arrangement for which the backward arcs make up a feedback arc set. For more details see~\cite{di1999graph,handbook}. Using the implementations of~\cite{simpson2016efficient}, GreedyFAS runs very fast, in $O(n+m)$ time, and is suitable for their extra large webgraphs.
%the Algorithm~\ref{alg:greedyfas}.
The pseudocode for GreedyFAS, as described in~\cite{di1999graph} and~\cite{simpson2016efficient}, is presented in Algorithm~\ref{alg:greedyfas}.
\begin{algorithm}
\caption{GreedyFAS}
\label{alg:greedyfas}
\begin{algorithmic}[0]
    \State \textbf{Input:} Directed graph $G=(V, E)$
    \State \textbf{Output:} Linear Arrangement A
    \State $s_1\gets\emptyset$, $s_2\gets\emptyset$
    \While{$G\not=\emptyset$}
        \While{$G$ contains a sink}
            \State choose a sink $u$
            \State $s_2\gets u s_2$
            \State $G\gets G \backslash u$
        \EndWhile
        \While{$G$ contains a source}
            \State choose a source $u$
            \State $s_1\gets s_1 u$
            \State $G\gets G \backslash u$
        \EndWhile
        \State choose a node $u$ for which $\delta(u)$ is a maximum
        \State $s_1\gets s_1 u$
        \State $G\gets G \backslash u$
    \EndWhile
    \State \textbf{return} $s=s_1 s_2$
\end{algorithmic}

\end{algorithm}
\subsection{SortFAS}
The SortFAS algorithm was introduced in 2011 by Brandenburg \& Hanauer~\cite{brandenburg2011sorting}. The algorithm is an extension of the KwikSortFAS heuristic by Ailon et al.~\cite{ailon2008aggregating}, which is an approximation algorithm for the FAS problem on tournaments. With SortFAS, Brandenburg \& Hanauer extended the above heuristic to work for general directed graphs. It uses the underlying idea that the nodes of a graph can be sorted into a desirable Linear Arrangement based on the number of back arcs induced.

In the case of SortFAS, the nodes are processed in order of their ordering $(v_1 ... v_n)$. The algorithm goes through $n$ iterations. In the the $i$-th iteration, node $v_i$ is inserted into the linear arrangement in the best position based on the first $i-1$ nodes which are already placed. The best position is the one with the least number of back arcs induced by $v_i$. In case of a tie the leftmost position is taken. Using the implementation of~\cite{simpson2016efficient}, SortFAS runs in $O(n^2)$ time. The pseudocode for SortFAS, as described in~\cite{simpson2016efficient}, is presented in Algorithm~\ref{alg:sortfas}.

\begin{algorithm}
\caption{SortFAS}
\label{alg:sortfas}
\begin{algorithmic}[0]
    \State \textbf{Input:} Linear arrangement A
    \For{\textbf{each} node $v$ in $A$}
        \State $val\gets 0$, $min\gets0$, $loc\gets$ position of $v$
        \For{\textbf{each} position $j$ from $loc-1$ down to $-$}
            \State $w\gets$ node at position $j$
            \If{arc $(v, w)$ exists}
                \State $val\gets val-1$
            \ElsIf{arc $(w, v)$ exists}
                \State $val\gets val+1$
            \EndIf
            \If{$val\leq min$}
                \State $min\gets val, loc\gets j$
            \EndIf
        \EndFor
        \State insert $v$ at position $loc$
    \EndFor
\end{algorithmic}
\end{algorithm}

\section{Our proposed Approach}
\label{se:newAlgorithm}
Our approach is based on running the well known PageRank algorithm~\cite{DBLP:journals/cn/BrinP98,page1999pagerank} on the directed line digraph of the original directed graph.
The \emph{line graph} of an undirected graph $G$ is another graph $L(G)$ that is constructed as follows: each edge in $G$ corresponds to a node in $L(G)$ and for every two edges in $G$ that are adjacent to a node $v$ an edge is placed in $L(G)$ between the corresponding nodes. Clearly, the number of nodes of a line graph is $m$ and the number of edges is proportional to the sum of squares of the degrees of the nodes in $G$, see~\cite{pemmaraju_skiena_2003}.
% \begin{equation}
% |E|_{undirected} = \frac{1}{2}\sum_{i=1}^n d(v_i)^2-m
% \end{equation}
If $G$ is a directed graph, its \emph{directed line graph} (or \emph{line digraph}) $L(G)$ has one node for each edge of $G$. 
%Two nodes representing directed edges from $u$ to $v$ (outgoing from $v$) and from $v$ to $w$,
Two nodes representing directed edges incident upon $v$ in $G$ (one incoming into $v$, and one outgoing from $v$), called $L(u,v)$, and $L(v,w)$, are connected by a directed edge from $L(u,v)$ to $L(v,w)$ in $L(G)$.  In other words, every edge in $L(G)$ represents a directed path in $G$ of length two.  Similarly, the number of nodes of a line digraph is $m$ and the number of edges is proportional to $\sum_{u\in V} [d^+(u) \times d^-(u)])$
% \begin{equation}
% |E|_{directed} = \frac{1}{2}\sum_{i=1}^n d^-(v_i)*d^+(v_i)-m
% \end{equation}
%sum of  $[d^+(u) \times d^-(u)]$ over all $u$ in $G$ $\sum$.  
Hence, the size of $L(G)$ is $O(m + \sum_{u\in V} [d^+(u) \times d^-(u)])$.  
%where $d^+(u)$ denotes the out-degree of $u$ and $d^-(u)$ denotes the in-degree of $u$.
%... , see (references here).

Given a digraph $G=(V,E)$ our approach is to compute its line digraph, $L(G)$, run a number of iterations of PageRank on $L(G)$ and remove the node of highest PageRank in $L(G)$.  Our experimental results indicate that PageRank values converge reasonably well within five iterations.

A digraph $G$ is \emph{strongly connected} if for every pair of vertices of $G$ there is a cycle that contains them. If $G$ is not strongly connected, it can be decomposed into its \emph{strongly connected components (SCC)} in linear time~\cite{TarjanSCC}. An SCC of $G$ is a subgraph that is strongly connected, and is maximal, in the sense that no additional edges or vertices of $G$ can be included in the subgraph without breaking its property of being strongly connected. If each SCC is contracted to a single vertex, the resulting graph is a directed acyclic graph (DAG).  It follows that feedback arcs can exist only within some (SCC) of $G$.  Hence we can apply this approach inside each SCC, using their corresponding line digraph, and remove the appropriate edges from each SCC.  This approach will avoid performing several useless computations and thus reduce the running time of the algorithm.

\subsection{Line Graph}
In order to obtain the line digraph of G, we use a DFS-based approach. First, for each edge $(u, v)$ of G, we create a node $(u, v)$ in $L(G)$ and then run the following recursive procedure. For a node $v$, we mark it as visited and iterate through each one of its outgoing edges. For each outgoing edge $(v, u)$ of $v$, we add an edge in $L(G)$ from the $prev$ $L(G)$ node that was processed before the procedure's call to the node $(v, u)$. Afterwards we call the same procedure for $u$ if it's not visited with $(v, u)$ as $prev$. If $u$ is visited we add an edge from $(v, u)$ to each one of $L(G)$'s nodes corresponding from $u$. Since this technique is based on DFS, the running time is $O(n+m+|L(G)|)$. The pseudocode for computing a line digraph is presented in Algorithm~\ref{algo:linedigraph}.

\begin{algorithm}
\caption{LineDigraph}
\label{algo:linedigraph}
\begin{algorithmic}[0]
    \State \textbf{Input:} Digraph $G=(V, E)$
    \State \textbf{Output:} Line Digraph $L(G)$ of $G$
    \State Create a line digraph $L(G)$ with every edge of $G$ as a node
    \State $v\gets$ random node of $G$
    \State \hfill
    \Procedure{GetLineGraph}{$G, L(G), v, prev$}
        \State mark $v$ as $visited$
        \For{\textbf{each} edge $e=(v, u)$ outgoing of $v$}
            \State $z\gets$ node of $L(G)$ representing $e$
            \State create an edge in $L(G)$ from $prev$ to $z$\Comment{Given that $prev$ is not nill}
            \If{$u$ is not $visited$}
                \State GetLineGraph$(G, L(G), u, z)$
            \Else
                \For{\textbf{each} node $k$ in $L(G)$ that originates from $u$}
                    \State create an edge in $L(G)$ from $z$ to $k$
                \EndFor
            \EndIf
        \EndFor
    \EndProcedure
\end{algorithmic}
\end{algorithm}

\subsection{PageRank}
PageRank was first introduced by Brin \& Page in 1998~\cite{DBLP:journals/cn/BrinP98,page1999pagerank}. It was developed in order to determine a measure of importance of web pages in a hyperlinked network of web pages. The basic idea is that PageRank will assign a score of importance to every node (web page) in the network. The underlying assumption is that important nodes are those that receive many ``recommendations'' (in-links) from other important nodes (web pages). In other words, it is a link analysis algorithm that assigns numerical scores to the nodes of a graph in order to measure the importance of each node in the graph. PageRank works by counting the number and quality/importance of edges pointing to a node and then estimate the importance of that node.  We use a similar approach in order to determine the importance of edges in a directed graph. The underlying assumption of our technique is that the number of cycles that contain a specific edge $e$ will be reflected in PageRank score of $e$. Thus the removal of edges with high PageRank score is likely to break the most cycles in the graph.

Given a graph with  $n$ nodes and $m$ edges, PageRank starts by assigning an initial score of $1/n$ to all the nodes of a graph. Then for a predefined number of iterations each node divides its current score equally amongst its outgoing edges and then passes these values to the nodes it is pointing to. If a node has no outgoing links then it keeps its score to itself. Afterwards, each node updates its new score to be the sum of the incoming values. It is obvious that after enough iterations all PageRank values will inevitably gather in the sinks of the graph. In use cases where that is a problem a damping factor is used, where each node gets a percentage of its designated score and the rest gets passed to all other nodes of the graph. For our use case we have no need for this damping factor as we want the scores of the nodes to truly reflect their importance. The number of iterations depends on the size and structure of a graph. We found that for small and medium graphs, which is the case in the scenario for graph visualization, about five iterations were enough for the scores of the nodes to converge. Depending on the implementation, PageRank can run in $O(k(n+m))$ time, where $k$ is the number of iterations.
%, $n$ is the number of nodes and $m$ is the number of edges. 
The pseudocode for PageRank is presented in Algorithm~\ref{algo:pagerank}.

\begin{algorithm}
\caption{PageRank}
\label{algo:pagerank}
\begin{algorithmic}[0]
    \State \textbf{Input:} Digraph $G=(V, E)$, number of iterations $k$
    \State \textbf{Output:} PageRank scores of $G$
    \For{\textbf{each} node $v$ in $G$}
        \State $PR(v)\gets\frac{1}{|V|}$
    \EndFor
    \For{$k$ iterations}
        \For{\textbf{each} node $v$ in $G$}
            \State $PR(v)\gets \sum_{u\in in(v)}\frac{PR_{old}(u)}{|out(u)|}$
        \EndFor
    \EndFor
    \State \textbf{return} $PR$
\end{algorithmic}
\end{algorithm}

\subsection{PageRankFAS}
% afto mporei na xreiazetai rephrase to intro
Our proposed algorithm is based on the concepts of PageRank and Line Digraphs. The idea behind \textit{PageRankFAS} is that we can score the edges of $G$ based on their involvement in cycles: For each strongly connected component ($s_1, s_2, ..., s_j$) of $G$, it computes the line digraph $L(s_i)$ of the $i$-th strongly connected component, to transform edges to nodes; next it runs the PageRank algorithm on $L(s_i)$ to obtain a score for each edge of $s_i$ in $G$.

We observed that the nodes of the line digraphs with the highest PageRank score correspond to edges that are involved in the most cycles of $G$. We also observed that the nodes of the line digraphs with lower score correspond to edges of $G$ with low involvement in cycles. Using this knowledge, we run PageRankFAS for a number of iterations. In each iteration, we use PageRank to calculate the node scores of each $L(s_i)$ and remove the node(s) with the highest PageRank score, also removing the corresponding edge(s) from $G$. We repeat this process until $G$ becomes acyclic. The pseudocode is presented in Algorithm~\ref{algo:pagerankFAS}.

\begin{algorithm}
\caption{PageRankFAS}
\label{algo:pagerankFAS}
%\end{algorithm}
\begin{algorithmic}[0]
    \State \textbf{Input:} Digraph $G=(V, E)$
    \State \textbf{Output:} Feedback Arc Set of $G$
    \State $fas\gets\emptyset$
    \While{$G$ has cycles}
        \State Let ($s_1, s_2, ..., s_j$) be the strongly connected components of $G$
        \For{\textbf{each} strongly connected component $s_i$}
            \State Create a line digraph $L(s_i)$ with every edge of $s_i$ as a node
            \State $v\gets$ random node of $s_i$
            \State GetLineGraph($s_i, L(s_i), v, nill$)
            \State PageRank($L(s_i)$)
            \State $u\gets$ node of $L(s_i)$ with highest PageRank value
            \State $e\gets$ edge of $G$ corresponding to $u$
            \State Add $e$ to $fas$
            \State Remove $e$ from $G$
        \EndFor
    \EndWhile
    \State \textbf{return} $fas$
\end{algorithmic}
\end{algorithm}

\section{Experiments and Discussion}
\label{se:experiments}
Here we report the experimental results and describe some details of our setup. All of our algorithms are implemented in Java 8 using the WebGraph framework~\cite{BRSLLP,BoVWFI} and tested on a 
single machine with Apple's M1 processor, 
8GB of RAM and running macOS Monterey 12.

\hfill\\
\noindent \textbf{Datasets:} In order to evaluate our proposed heuristic algorithm we used four different datasets:
\begin{enumerate}
  \item Randomly generated graphs with 100, 200, 400, 1000, 2000, 4000 nodes and an average out-degree of 1.5, 3 and 5 each.
  \item Three directed graphs from the datasets in graphdrawing.org%\footnote{\label{graphdrawingorg}\url{http://graphdrawing.org}},
   , suitably modified in order to contain cycles (since the originals are DAGs).
  \item Randomly generated graphs with 50, 100 and 150 nodes and average out-degrees of 1.5, 3, 5, 8, 10 and 15 each.
  \item Two webgraphs from the Laboratory of Web Algorithmics\footnote{\url{https://law.di.unimi.it/datasets.php}}, also used in~\cite{simpson2016efficient}.
\end{enumerate}

We randomly generate a total of 36 graphs using a predefined number of nodes, average out-degree and back edge percentage, and we repeat the process 10 times.
%The total number of edges is based on the number of nodes and average out-degree. 
%Next, we add the predefined percentage of edges as back edges and the rest as forward edges. Finally we shuffle the node IDs in order to prevent any bias in the traversal of the graph.
By construction, this model has the advantage that we know in advance an upper bound to the FAS size, since the number of randomly created back edges divided by the total number of edges, is an upper-bound to the size of a minimum FAS. Finally, in order to avoid having abrupt results due to randomness, for each case we run the three algorithms on 10 created graphs and report the average numbers. This smooths out several points in our curves.

\subsection{FAS with Respect to the Number of Nodes}

\begin{figure}[p]
    \centering
    \begin{minipage}{0.46\columnwidth}
        \includegraphics[width=\linewidth]{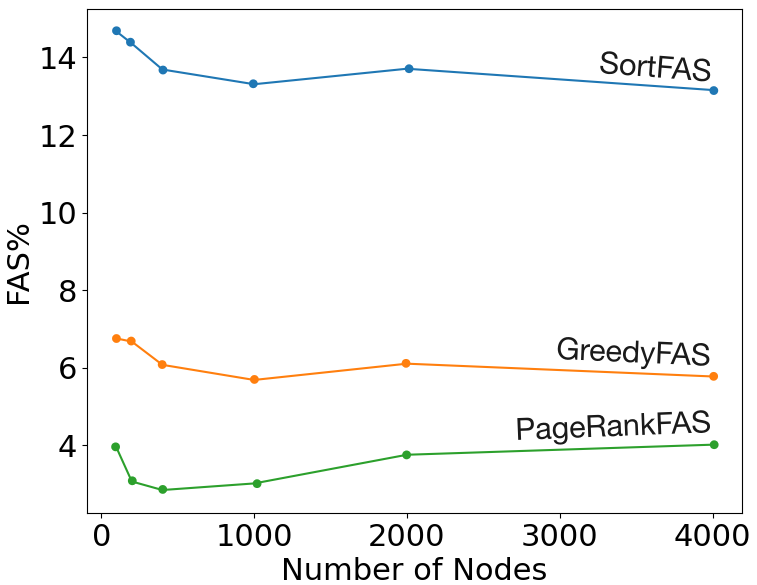}
        \caption*{\textbf{(a)} Graphs with average out-degree 1.5}
        \hfill\\
        \includegraphics[width=\linewidth]{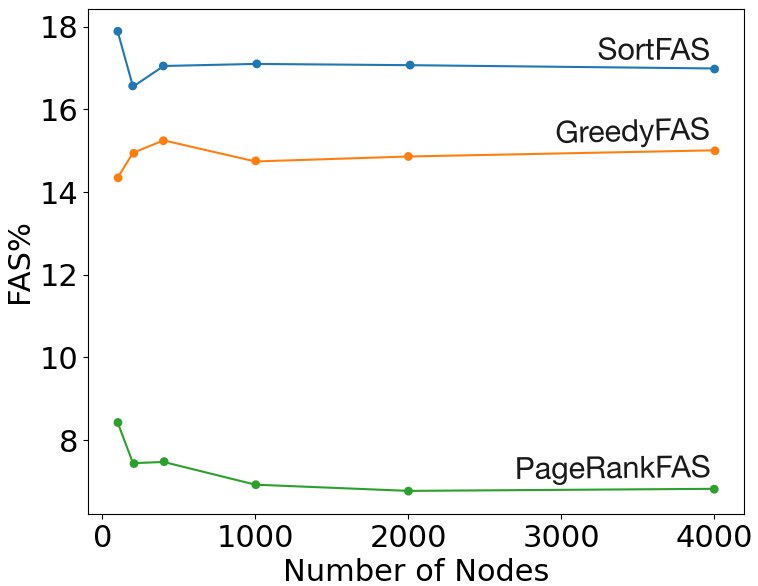}
        \caption*{\textbf{(b)} Graphs with average out-degree 3}
        \hfill\\
        \includegraphics[width=\linewidth]{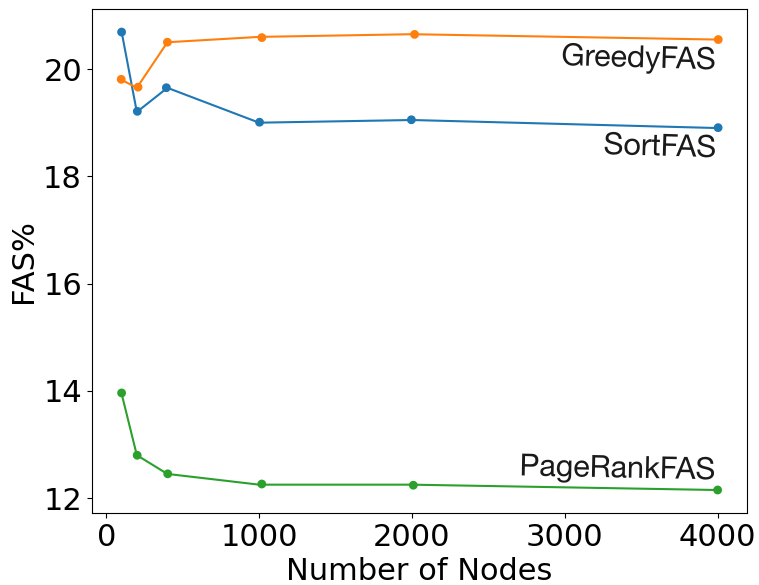}
        \caption*{\textbf{(c)} Graphs with average out-degree 5}
        \caption{FAS percentage for graphs with increasing number of nodes and three different average out-degrees.}
        \label{fig:fas-comparative}
    \end{minipage}\hfill
    \begin{minipage}{0.46\columnwidth}
        \includegraphics[width=\linewidth]{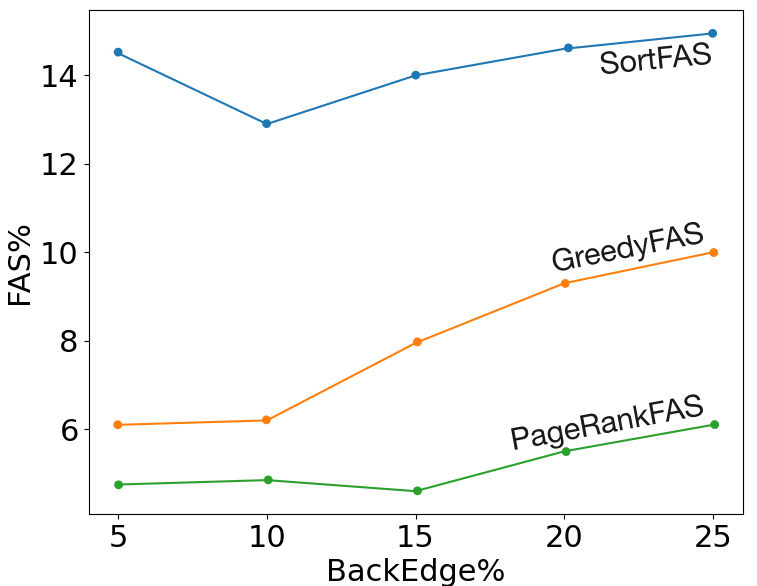}
        \caption*{\textbf{(a)} Graph with 50 nodes and 75 edges before modification}
        \hfill\\
        \includegraphics[width=\linewidth]{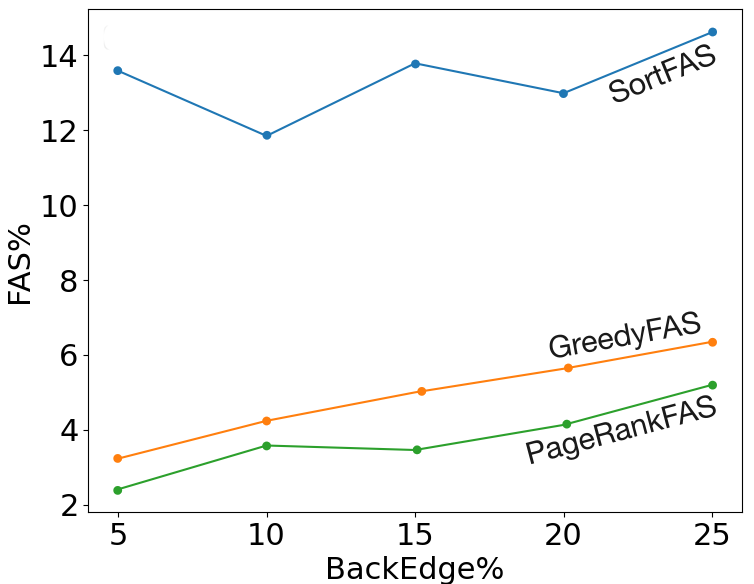}
        \caption*{\textbf{(b)} Graph with 75 nodes and 86 edges before modification}
        \hfill\\
        \includegraphics[width=\linewidth]{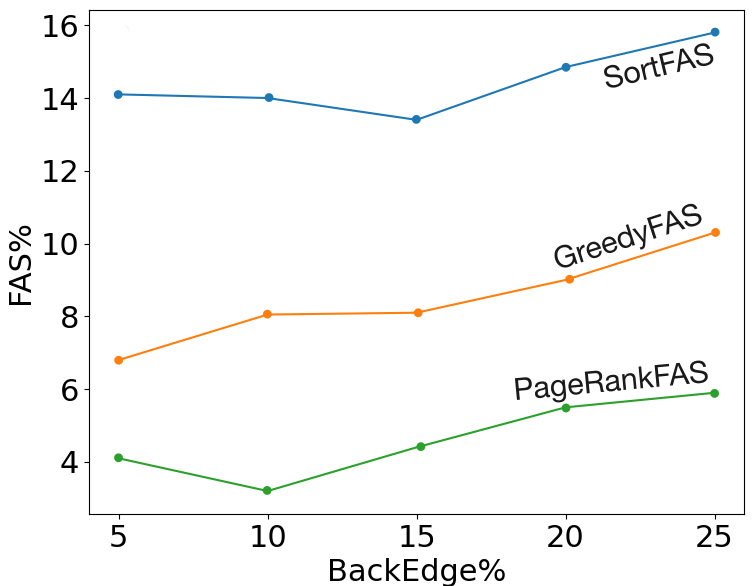}
        \caption*{\textbf{(c)} Graph with 99 nodes and 154 edges before modification}
        \caption{FAS percentage for 3 types of graphs from graphdrawing.org and for various numbers of back edges.}
        \label{fig:fas-back}
    \end{minipage}
\end{figure}

The first set of experiments gives us an idea of how PageRankFAS performs on graphs, with varying number of nodes in comparison to the other two algorithms.
It is noteworthy that in most cases the FAS found by PageRankFAS is less than 50\% of the FAS found by GreedyFAS and SortFAS. As a matter of fact, for large visualization graphs with 4,000 nodes and 12,000 edges the reduction in the FAS size is almost 55\% with respect to the FAS produced by GreedyFAS. The execution time taken by PageRankFAS is less than one second for graphs up to 1,000 nodes, which is similar to the time of the other two heuristics. For the larger graphs, even up to 4,000 nodes the time required is less than 8 seconds, whereas, the other heuristics run in about 1-2 seconds. The results of this experiment are shown in Figure~\ref{fig:fas-comparative}. It is interesting to note that the performance of SortFAS is better than the performance of GreedyFAS as the graphs become denser, and in fact, SortFAS actually out-performs GreedyFAS  when the graphs have an average out-degree 5 and above, see Figure~\ref{fig:fas-comparative}(c). 
%"It is interesting to note that the performance of SortFAS improves as the graphs become denser”

\subsection{FAS with Respect to The Number of Back Edges}
The second type of experiments make use of three graphs from graphdrawing.org. 
Since these graphs 
are directed acyclic, we randomly added back edges in different percentages of the total number of edges.  We did this in a controlled manner in order to know in advance an upper bound of FAS. PageRankFAS gave by far the best FAS results and GreedyFAS also produced FAS with sizes mostly below 10\%. SortFAS was not competitive in this dataset. The results are shown in Figure~\ref{fig:fas-back}. The execution time taken by PageRankFAS is well below $0.15$ of a second for all graphs, which is similar to the other two heuristics.

\subsection{FAS with Respect to the Average Out-Degree}
\begin{figure}[p]
    \centering
    \begin{minipage}{0.46\columnwidth}
        \includegraphics[width=\linewidth]{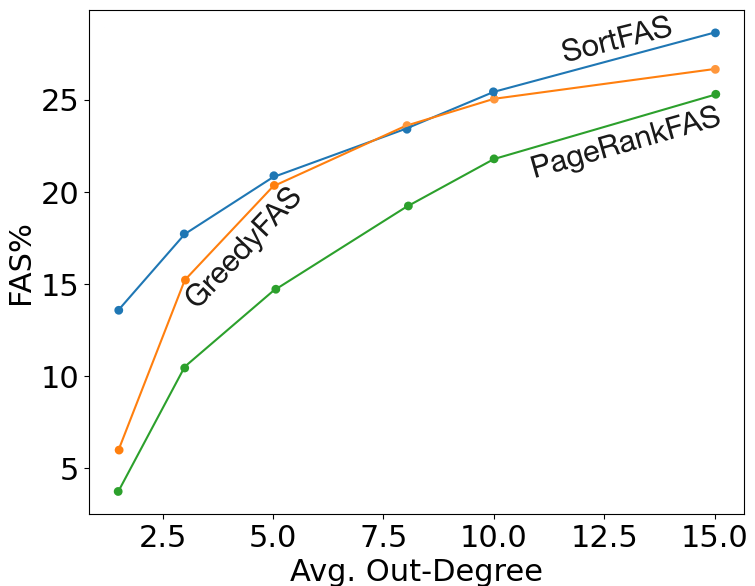}
        \caption*{\textbf{(a)} Graphs with 50 nodes}
        \hfill\\
        \includegraphics[width=\linewidth]{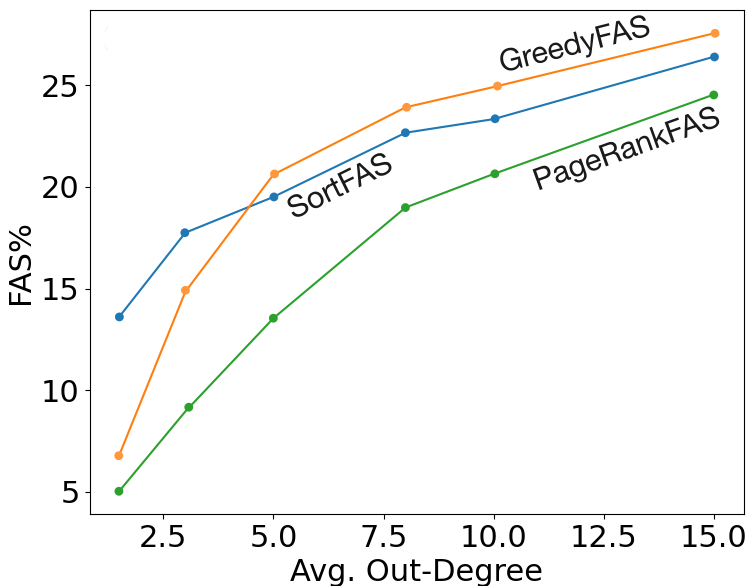}
        \caption*{\textbf{(b)} Graphs with 100 nodes}
        \hfill\\
        \includegraphics[width=\linewidth]{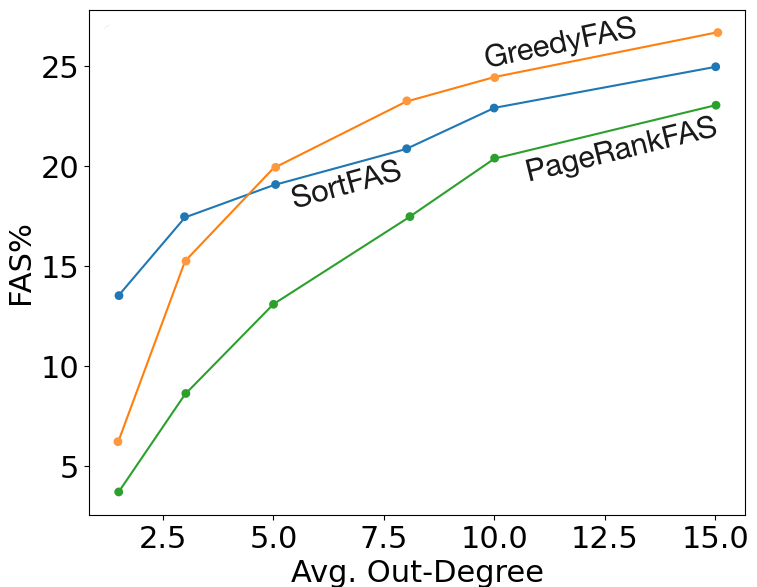}
        \caption*{\textbf{(c)} Graphs with 150 nodes}
        \caption{FAS percentage depending on the average out-degree of three different types of graphs.}
        \label{fig:fas-density}
    \end{minipage}\hfill
    \begin{minipage}{0.46\columnwidth}
        \includegraphics[width=\linewidth]{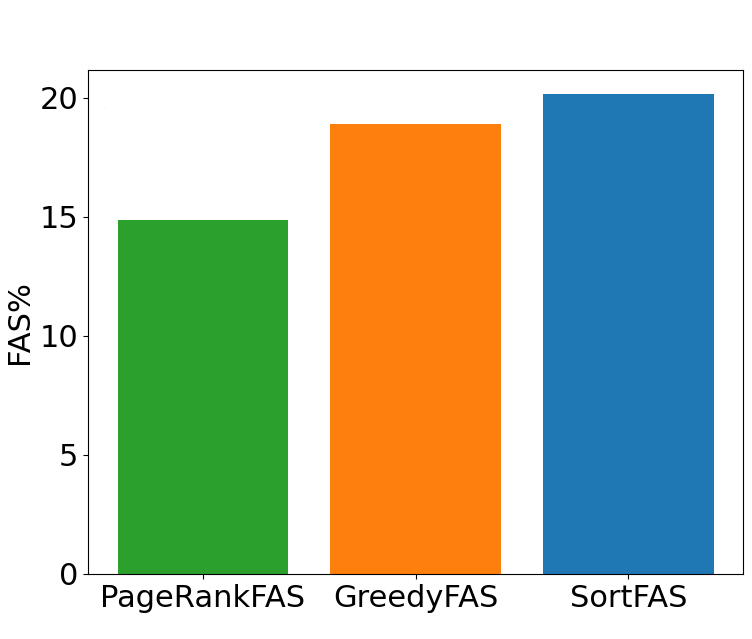}
        \caption*{\textbf{(a)} wordassociation-2011}
        \includegraphics[width=\linewidth]{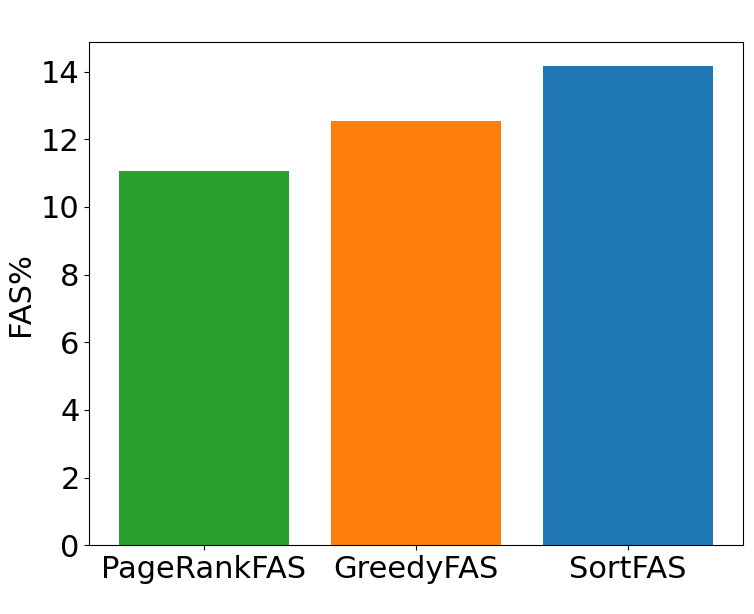}
        \caption*{\textbf{(b)} enron}
        \caption{FAS percentage on two webgraphs.}
        \label{fig:fas-webgraphs}
        \includegraphics[width=\linewidth]{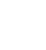}
    \end{minipage}
\end{figure}

Motivated by the results shown in Figure~\ref{fig:fas-comparative}(c) we decided to investigate the correlation between the density of a graph and its potential FAS percentage. In this experiment, we created 18 different graphs, six of them with 50 nodes, six with 100 nodes and six with 150 nodes as follows: For each node size (i.e., 50, 100, 150) six graphs with average out-degrees 1.5, 3, 5, 8, 10 and 15. Again, as with our previous experiments, the results reported here are the averages of 10 runs in order to compensate for the randomness of each graph and to get smoother curves. The results of this experiment are shown in Figure~\ref{fig:fas-density}.

The results of PageRankFAS are consistently better than the results of GreedyFAS and SortFAS for all graphs. The results of GreedyFAS and SortFAS are very close to each other, for the graphs with 50 nodes. Notice however that, SortFAS outperforms GreedyFAS when the number of nodes exceeds 100 and the average out-degree exceeds five. This is aligned with the results shown in Figure~\ref{fig:fas-comparative}(c).
Furthermore, as expected, when the average out-degree increases the FAS size clearly increases. Consequently, all techniques seem to converge at higher percentages of FAS size.
Again, PageRankFAS runs in a small fraction of a second for all graphs, which is similar to the running times of the other two heuristics.

\subsection{PageRankFAS on Webgraphs}
The experiments reported in~\cite{simpson2016efficient} use large and extra large benchmark webgraphs.  Their smaller benchmarks are \textit{wordassociation-2011} (with 10,617 nodes,  72,172 edges, which implies an average degree 6.80) and \textit{enron} (with 69,244 nodes, 276,143 edges, which implies an average degree 3.86).  

The authors report that the sizes of a FAS found by GreedyFAS and SortFAS for wordassociation-2011 are 18.89\% and 20.17\%, respectively~\cite{simpson2016efficient}. We ran PageRankFAS for wordassociation-2011 and obtained a FAS of size 14.85\%.  Similarly, for webgraph enron they report a FAS of 12.54\% and 14.16\% respectively.  We ran PageRankFAS on webgraph enron and obtained a FAS of size 11.05\%.  The results are shown in Figure~\ref{fig:fas-webgraphs}.  As expected, and consistent with our experimental observations of the previous subsections, the FAS size of the denser webgraph (wordassociation-2011) is larger than the FAS size of the sparser graph (enron), as computed by all heuristics.

Unfortunately, the required execution time of our algorithm does not allow us to test it on the larger webgraphs used in~\cite{simpson2016efficient}. However, it is interesting that there exists a FAS of smaller size for these large graphs, which, to the best of our knowledge, was not known before.

\newpage
\section{Conclusions}
\label{se:Conclusions}
We presented a heuristic algorithm for computing a FAS of minimum size based on PageRank. Our experimental results show that the size of a FAS computed by our heuristic algorithm is typically about 50\% smaller than the sizes obtained by the best previous heuristics. Our algorithm is more time consuming than the best previous heuristics, but it's running time is reasonable for graphs up to 4,000 nodes. For smaller graphs, up to 1,000 nodes, the execution time is well below one second, which is similar to the running times of the other two heuristics. Therefore, this is acceptable for graph drawing applications. An interesting side result is that we found out that the FAS-size of two large graphs is significantly less than it was known before.  Since it is NP-hard to compute the minimum FAS, the optimum solution for these webgraphs is unknown.  Hence, we do not know how close our solutions are to the optimum.
It would be interesting to investigate techniques to speedup PageRankFAS in order to make it more applicable to larger webgraphs.
%
% ---- Bibliography ----
%

\bibliographystyle{splncs04}
\clearpage
\bibliography{main} 

\begin{thebibliography}{10}
\providecommand{\url}[1]{\texttt{#1}}
\providecommand{\urlprefix}{URL }
\providecommand{\doi}[1]{https://doi.org/#1}

\bibitem{ailon2008aggregating}
Ailon, N., Charikar, M., Newman, A.: Aggregating inconsistent information:
  ranking and clustering. Journal of the ACM (JACM)  \textbf{55}(5),  1--27
  (2008). \doi{10.1145/1411509.1411513}

\bibitem{BRSLLP}
Boldi, P., Rosa, M., Santini, M., Vigna, S.: Layered label propagation: A
  multiresolution coordinate-free ordering for compressing social networks. In:
  Srinivasan, S., Ramamritham, K., Kumar, A., Ravindra, M.P., Bertino, E.,
  Kumar, R. (eds.) Proceedings of the 20th international conference on World
  Wide Web. pp. 587--596. ACM Press (2011)

\bibitem{BoVWFI}
Boldi, P., Vigna, S.: The {W}eb{G}raph framework {I}: {C}ompression techniques.
  In: Proc. of the Thirteenth International World Wide Web Conference (WWW
  2004). pp. 595--601. ACM Press, Manhattan, USA (2004)

\bibitem{brandenburg2011sorting}
Brandenburg, F.J., Hanauer, K.: Sorting heuristics for the feedback arc set
  problem. In: Technical Report MIP-1104. University of Passau Germany (2011)

\bibitem{DBLP:journals/cn/BrinP98}
Brin, S., Page, L.: The anatomy of a large-scale hypertextual web search
  engine. Comput. Networks  \textbf{30}(1-7),  107--117 (1998)

\bibitem{budak2011limiting}
Budak, C., Agrawal, D., El~Abbadi, A.: Limiting the spread of misinformation in
  social networks. In: Proceedings of the 20th international conference on
  World wide web. pp. 665--674 (2011). \doi{10.1145/1963405.1963499}

\bibitem{di1999graph}
Di~Battista, G., Eades, P., Tamassia, R., Tollis, I.G.: Graph drawing,
  vol.~357. Prentice Hall, Upper Saddle River, NJ (1999)

\bibitem{eades1993fast}
Eades, P., Lin, X., Smyth, W.F.: A fast and effective heuristic for the
  feedback arc set problem. Information Processing Letters  \textbf{47}(6),
  319--323 (1993)

\bibitem{he2012influence}
He, X., Song, G., Chen, W., Jiang, Q.: Influence blocking maximization in
  social networks under the competitive linear threshold model. In: Proceedings
  of the 2012 siam international conference on data mining. pp. 463--474. SIAM
  (2012)

\bibitem{johnson1982np}
Johnson, D.S.: The np-completeness column: An ongoing gulde. Journal of
  Algorithms  \textbf{3}(4),  381--395 (1982)

\bibitem{karp1972reducibility}
Karp, R.M.: Reducibility among combinatorial problems. In: Complexity of
  computer computations, pp. 85--103. Springer (1972)

\bibitem{handbook}
Nikolov, N.S., Healy, P.: Hierarchical Drawing Algorithms, in Handbook of Graph
  Drawing and Visualization, ed. Roberto Tamassia. CRC Press (2014), pp.
  409-453

\bibitem{JGAA-502}
{Ortali}, G., {Tollis}, I.G.: A new framework for hierarchical drawings.
  Journal of Graph Algorithms and Applications  \textbf{23}(3),  553--578
  (2019). \doi{10.7155/jgaa.00502}

\bibitem{page1999pagerank}
Page, L., Brin, S., Motwani, R., Winograd, T.: The pagerank citation ranking:
  Bringing order to the web. Tech. rep., Stanford InfoLab (1999)

\bibitem{pemmaraju_skiena_2003}
Pemmaraju, S., Skiena, S.: Computational Discrete Mathematics: Combinatorics
  and Graph Theory with Mathematica ®. Cambridge University Press (1990)

\bibitem{simpson2016clearing}
Simpson, M., Srinivasan, V., Thomo, A.: Clearing contamination in large
  networks. IEEE Transactions on Knowledge and Data Engineering
  \textbf{28}(6),  1435--1448 (2016). \doi{10.1109/TKDE.2016.2525993}

\bibitem{simpson2016efficient}
Simpson, M., Srinivasan, V., Thomo, A.: Efficient computation of feedback arc
  set at web-scale. Proceedings of the VLDB Endowment  \textbf{10}(3),
  133--144 (2016). \doi{10.14778/3021924.3021930}

\bibitem{sugiyama1981methods}
Sugiyama, K., Tagawa, S., Toda, M.: Methods for visual understanding of
  hierarchical system structures. IEEE Transactions on Systems, Man, and
  Cybernetics  \textbf{11}(2),  109--125 (1981).
  \doi{10.1109/TSMC.1981.4308636}

\bibitem{TarjanSCC}
Tarjan, R.: Depth-first search and linear graph algorithms. SIAM Journal on
  Computing  \textbf{1}(2),  146--160 (1972)

\end{thebibliography}
\end{document}